\documentclass{aa} 

\usepackage{graphicx}
\usepackage{textcomp, gensymb}
\usepackage{booktabs}
\usepackage{tablefootnote}
\usepackage[flushleft]{threeparttable}
\usepackage{subcaption}
\usepackage{xcolor}

\usepackage[varg]{txfonts}

\usepackage[]{hyperref}
\hypersetup{
    colorlinks,
    citecolor=blue,
    filecolor=blue,
    linkcolor=blue,
    urlcolor=blue
}

\begin{document} 

   \title{Redshift $\sim$2.7 is not special}
  \subtitle{comment on ``Kolmogorov analysis of JWST deep survey galaxies''}

   \author{Prajwel Joseph\inst{1, *}}

   \institute{Indian Institute of Astrophysics, Bangalore 560034, India\\
              \email{prajwel.pj@gmail.com}
             }

   \date{Received January 15, 2000; accepted March 16, 2000}

\abstract
{
\citet{galikyan2025kolmogorov} reported a statistically significant change in galaxy spectral properties at redshift $z$ $\simeq$ 2.7 based on a Kolmogorov Stochasticity Parameter analysis of JWST spectroscopic data of galaxies. In this comment, we demonstrate that their result is critically driven by a single outlier in the dataset. This outlier arises from the use of a questionable redshift estimate for one spectrum. When the outlier is removed or the redshift is corrected, the claimed transition at $z$ $\simeq$ 2.7 disappears entirely. By independently reproducing their analysis, we demonstrate that the claimed feature is not a robust statistical signal, but an artefact of this anomalous data point.
}

 \keywords{galaxies: high-redshift}

   \maketitle

\section{Introduction}
\citealt{galikyan2025kolmogorov} (hereafter, G25) analysed JWST spectroscopic data of galaxies published by \cite{price2025uncover} to check for changes in the spectral properties of the sample galaxies with redshift.
They selected galaxy spectra with redshift ($z$) quality flags marked as either ``secure'' or ``solid'' in the dataset derived from the Ultradeep NIRSpec and NIRCam ObserVations before
the Epoch of Reionization (UNCOVER) Cycle 1 Treasury
survey \citep{bezanson2024jwst}.
They imposed an additional selection criterion that the spectral peak be close to 656 nm in the rest frame.
From this subset, the authors identified the observed-frame wavelengths ($\lambda_{\text{obs}}$) corresponding to the spectral peaks and converted them to the rest frame ($\lambda_{\text{rest}}$) using the redshift values estimated by \cite{price2025uncover}. They then grouped the data into small redshift bins and computed the Kolmogorov Stochasticity Parameter (KSP) for the set of rest-frame peak wavelengths within each bin. By examining how the KSP values varied with redshift, G25 reported a statistically significant change in spectral properties at $z$ $\simeq$ 2.7, claiming a confidence level exceeding 99\%. 

A sudden transition in galaxy spectral properties at $z$ $\simeq$ 2.7 is of significant astrophysical interest. Motivated by this claim, we revisited the JWST spectroscopic dataset to investigate the origin of the reported feature. As part of this effort, we reproduced the analysis carried out by G25. In doing so, we found that the reported change in spectral properties at $z$ $\simeq$ 2.7 arises entirely from the inclusion of a single outlier in their dataset. We detail our findings in the following section.

\begin{figure}
	\centering
	\includegraphics[width=\columnwidth]{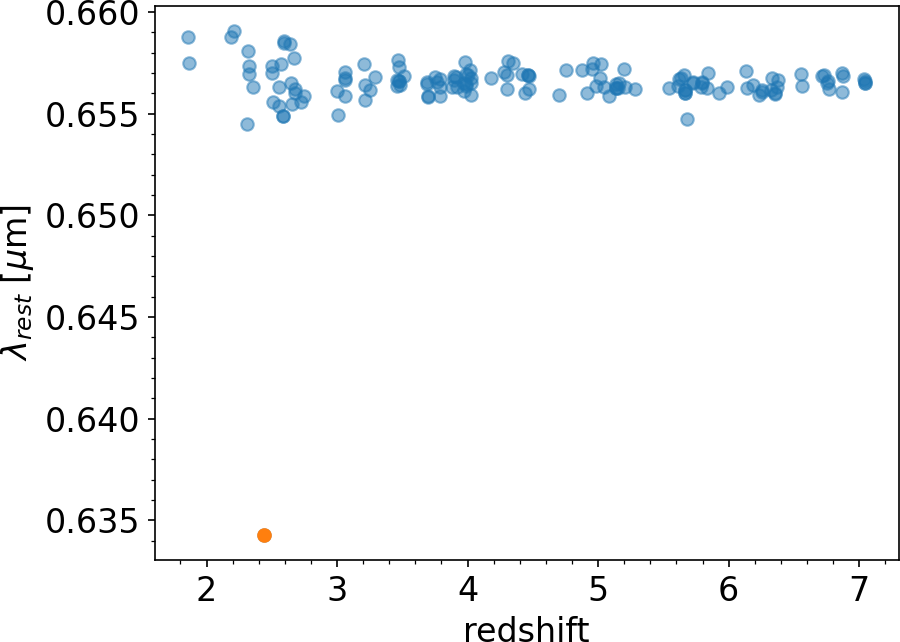}
	\caption{Rest-frame peak wavelengths near 656 nm in JWST galaxy spectra, as used by G25 in their analysis. The outlier, clearly offset from the rest of the distribution, is highlighted in orange.}
	\label{fig1}
\end{figure}

\section{Re-analysis of G25}

\begin{figure*}
    \centering
    \begin{subfigure}{0.36\textwidth}
        \includegraphics[width=\linewidth]{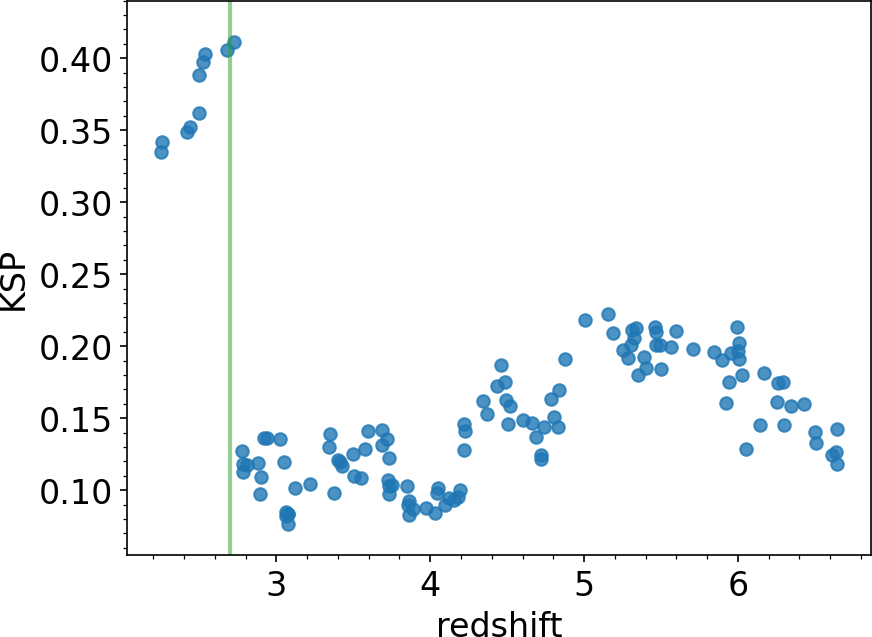}
        \caption{With outlier}
        \label{with_outlier}
    \end{subfigure}
    \hfill
    \begin{subfigure}{0.31\textwidth}
        \includegraphics[width=\linewidth]{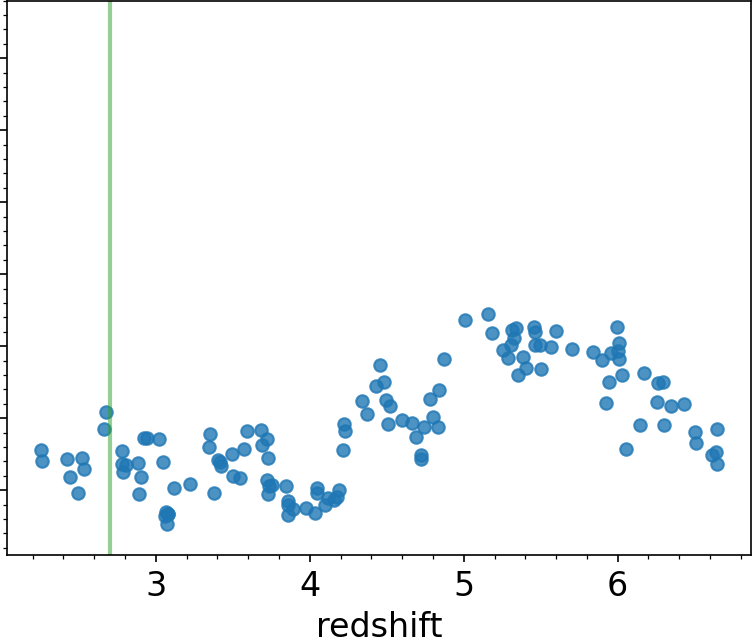}
        \caption{Without outlier}
        \label{without_outlier}
    \end{subfigure}
    \hfill
    \begin{subfigure}{0.31\textwidth}
        \includegraphics[width=\linewidth]{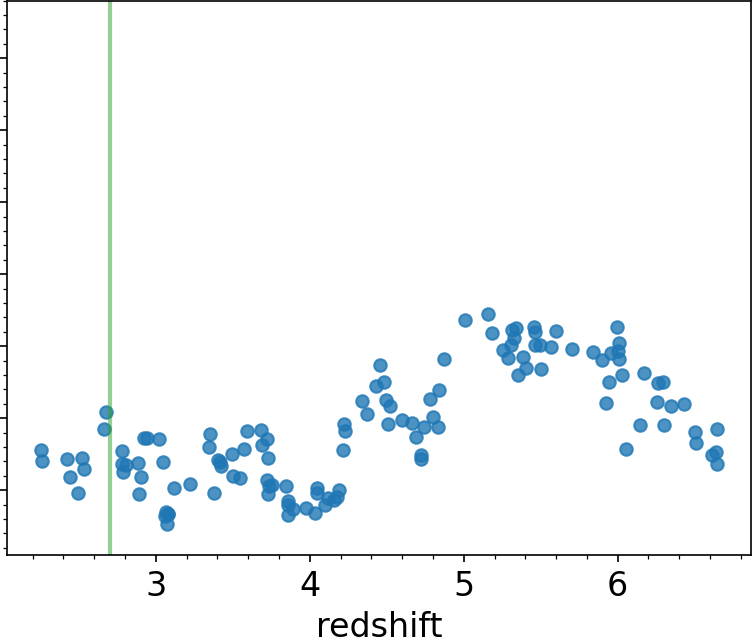}
        \caption{Outlier redshift corrected}
        \label{with_corrected_outlier}
    \end{subfigure}
    \caption{KSP versus redshift for the dataset used by G25 under different treatments of the outlier. (a) KSP computed with the outlier; (b) with the outlier excluded; (c) with the outlier redshift corrected using \texttt{z\_spec50}. In all panels, the vertical dashed line marks $z = 2.7$.}
    \label{fig:ksp_comparison}
\end{figure*}

\begin{figure}
	\centering
	\includegraphics[width=\columnwidth]{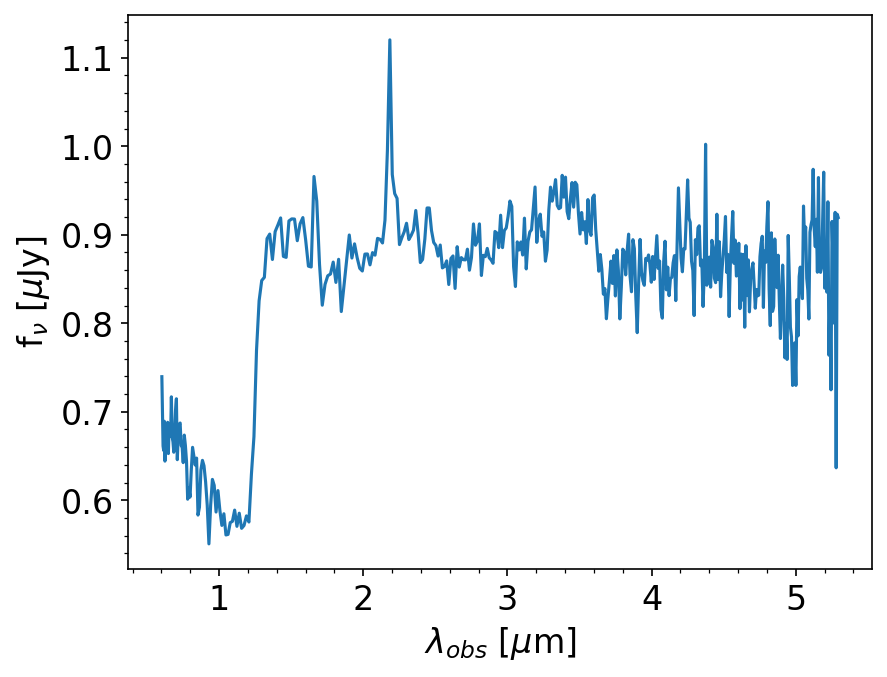}
	\caption{JWST spectrum corresponding to the outlier data point, shown in observed wavelengths. The redshifted H$\alpha$ emission line peak is clearly visible at $\sim$2.2~$\mu$m.}
	\label{outlier_spectrum}
\end{figure}

\begin{figure}
	\centering
	\includegraphics[width=\columnwidth]{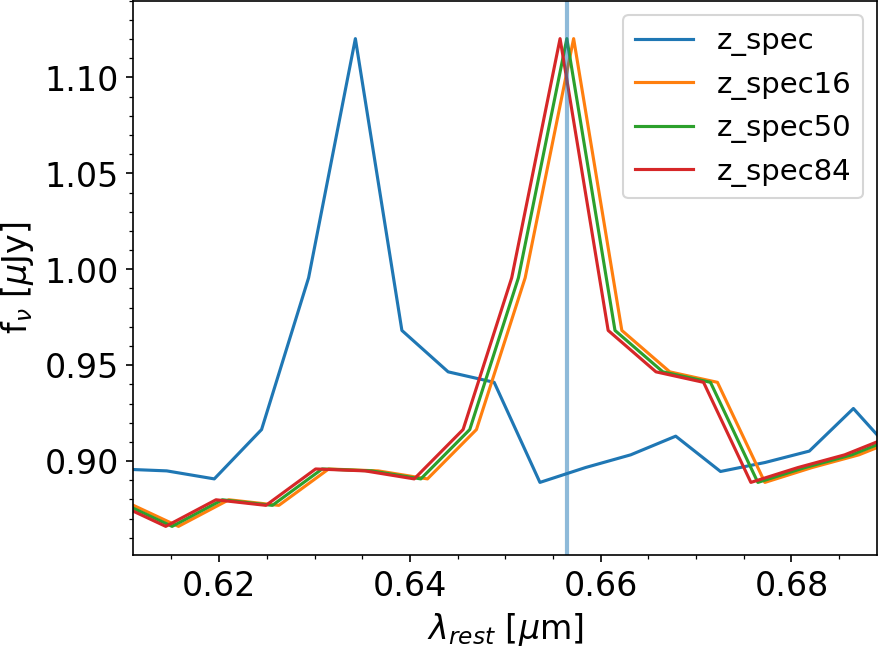}
	\caption{Region around the H$\alpha$ emission line in the spectrum corresponding to the outlier data point. 
    The observed wavelength ($\lambda_{\text{obs}}$) has been converted to the rest frame ($\lambda_{\text{rest}}$) using different redshift estimates: the blue curve uses \texttt{z\_spec}, while the orange, green, and red curves correspond to \texttt{z\_spec16}, \texttt{z\_spec50}, and \texttt{z\_spec84}, respectively. 
    The vertical line marks $\lambda_{\text{rest}}$ = 656.46~nm.}
	\label{outlier_halpha_peak}
\end{figure}

G25 computed KSP values using rest‑frame peak wavelengths in JWST galaxy spectra, specifically selecting peaks near 656 nm, where the H$\alpha$ emission line is typically prominent.
They interpreted KSP values thus computed as a measure of the degree of randomness in spectra and reported a statistically significant transition in the KSP distribution at $z \simeq 2.7$.
It is unclear how an analysis based solely on a single wavelength feature meaningfully captures the overall randomness of galaxy spectra. 
Additionally, G25 implicitly assume a uniform spectroscopic targeting strategy across all redshifts in the UNCOVER dataset. In practice, the source selection criteria varied significantly with redshift \citep{price2025uncover}, introducing redshift-dependent biases.
A further assumption underlying their analysis is that the H$\alpha$ emission line remains the dominant spectral feature across all redshifts. This is not generally valid, as galaxies at higher redshifts often exhibit stronger [O III] emission than H$\alpha$.
Nonetheless, accepting their methodology for the sake of argument, we have independently reproduced their analysis using the same UNCOVER sample.
Our re‑analysis demonstrates that the reported transition at $z \simeq 2.7$ arises entirely from the inclusion of a single outlier in their sample.

This outlier is visible in Figure 1 of G25 near $\sim$0.634~$\mu$m.
 We reproduce their figure following their described methodology (see Fig.~\ref{fig1}), highlighting the outlier in orange\footnote{The Python scripts used in this work can be accessed at \url{https://github.com/prajwel/no_special_redshift}.}.
Upon examining the corresponding data point ($\lambda_{obs}$ = 2.184 $\mu$m) in the publicly available dataset from \citet{price2025uncover}, we found that the relevant spectrum has four different redshift estimates: 2.444 (\texttt{z\_spec}), 2.324 (\texttt{z\_spec16}), 2.328 (\texttt{z\_spec50}), and 2.331  (\texttt{z\_spec84}).  
We note that the outlier arises only when using the \texttt{z\_spec} value to convert $\lambda_{\text{obs}}$. If any of the alternative redshift estimates are used, the corresponding rest-frame wavelength falls well within the general distribution of the other data points in Fig.~\ref{fig1}.
This suggests that the \texttt{z\_spec} value is likely a misestimate for the corresponding spectrum, and that one of the alternative values---remarkably consistent with one another---should be preferred. Moreover, visual inspection of the spectrum (see Fig.~\ref{outlier_spectrum}) confirms that the observed H$\alpha$ emission line lies close to the expected rest-frame wavelength when using any of the alternative redshift estimates (see Fig.~\ref{outlier_halpha_peak}).

It is important to note that wavelength calibration uncertainties in JWST/NIRSpec spectra, particularly in the low-resolution prism mode, can impact the accuracy of redshift estimates \citep{ferruit2022near, de2025rubies, d2025jades}. 
The prism mode suffers from reduced spectral resolution at shorter wavelengths ($R$ $\simeq$ 30 at $\sim$1~$\mu$m compared to $R$ $\simeq$ 300 at $\sim$5~$\mu$m), making it more difficult to obtain precise redshifts at lower $z$ \citep{jakobsen2022near}. 
In the case of the outlier spectrum discussed here, the large discrepancy in the \texttt{z\_spec} value relative to the other redshift estimates ($\sim$5\% difference), combined with visual confirmation of the H$\alpha$ line position, may point to an error likely exceeding the expected calibration uncertainties.

To demonstrate the influence of the outlier on the analysis by G25, we used a sliding window (bin), each containing 20 $\lambda_{\text{rest}}$ values sorted by redshift, and computed the KSP for the values in each bin.
We adopted the standard normal distribution as the reference theoretical distribution. The resulting plot of KSP versus redshift, based on the same data used by G25, is shown in Fig.~\ref{with_outlier}.

The Fig.~\ref{with_outlier} differs from Figure 2 of G25 in two main respects: the number of data points and the spread of KSP values. The difference in the number of points arises because G25 used 1000 randomly generated, overlapping redshift bins of varying sizes, while we used a uniform sliding bin of size 20. The difference in the spread of KSP values stems from their use of a generalised normal distribution with a free sharpness parameter fitted independently in each redshift bin. In contrast, we used a fixed standard normal distribution across all bins. Despite these methodological differences, our Fig.~\ref{with_outlier} successfully reproduces the sharp change in KSP around $z$ $\simeq$ 2.7 in Figure 2 of G25, along with other observed trends.

However, when the outlier is excluded from the analysis, this significant jump in KSP disappears entirely (see Fig.~\ref{without_outlier}). The same result holds when the outlier’s redshift is replaced with any of the alternative estimates discussed earlier (see Fig. \ref{with_corrected_outlier}). This demonstrates that the claimed change in KSP is entirely driven by a single inconsistent data point.

\section{Summary}

G25 analysed JWST spectra from \cite{price2025uncover}, selecting spectra with high-confidence redshift estimates and those with peaks near 656 nm in the rest frame. They examined how the degree of randomness in the rest-frame peak wavelengths varied with redshift, and reported a statistically significant change in spectral properties at $z$ $\simeq$ 2.7.

We show that their result is critically dependent on a single outlier in their dataset. This outlier, visible in their own Figure 1, arises from the choice of one specific redshift estimate out of several available for a given spectrum. When any of the alternative redshift estimates are used, the corresponding rest-frame wavelength no longer deviates from the overall distribution.

We independently reproduced the KSP versus redshift trend and confirmed that the reported feature at $z$ $\simeq$ 2.7 can be replicated---but only when the outlier is included. Once excluded or corrected, the jump in KSP disappears. This demonstrates that the claimed transition is not a robust statistical feature, but an artefact of a single inconsistent data point.

\begin{acknowledgements}
We thank the anonymous reviewer for their constructive comments and helpful suggestions, which have improved this article.
This work is based on a talk presented at a journal club meeting at the Indian Institute of Astrophysics.
The author is grateful to Koshy George and C. S. Stalin for insightful discussions.
Thanks are also due to Norayr Galikyan for responding to queries related to their work. 
Astropy, IPython, Matplotlib, NumPy, and 
Scipy were used for data analysis, 
viewing, and plotting \citep{astropy:2013, astropy:2018, astropy:2022, matplotlib, numpy, ipython2007, 2020SciPy-NMeth}.
This work is based on observations made with the NASA/ESA/CSA James Webb Space Telescope. The data were obtained from the Mikulski Archive for Space Telescopes at the Space Telescope Science Institute, which is operated by the Association of Universities for Research in Astronomy, Inc., under NASA contract NAS 5-03127 for JWST.
These observations are associated with JWST-GO-2561.
The specific observations analysed can be accessed via \url{https://dx.doi.org/10.17909/8k5c-xr27}.
\end{acknowledgements}

\bibliographystyle{aa}
\bibliography{references}

\end{document}